\documentclass[aps,prb,onecolumn,showpacs,floatfix]{revtex4}
\usepackage{graphicx}
\usepackage{amsmath}

\begin{document}

\title{Effect of spin-orbit coupling on magnetic and orbital order in MgV$_2$O$_4$}
\author{Ramandeep Kaur, T. Maitra  and  T. Nautiyal}
\affiliation{Department of Physics, Indian Institute of Technology Roorkee, Roorkee- 247667,
Uttarakhand, India}
\pacs {71.20.-b, 75.25.Dk, 71.70.Ej, 71.27.+a}
\date{\today}

\begin{abstract}
Recent measurements on MgV$_2$O$_4$ single crystal have reignited
the debate on the role of spin-orbit (SO) coupling in dictating the orbital order
in Vanadium spinel systems. Density functional theory calculations were performed using the full-potential linearized augmented-plane-wave method within the local spin density
approximation (LSDA), Coulomb correlated LSDA+U, and with SO interaction (LSDA+U+SO) to study the magnetic and orbital ordering in low temperature
phase of MgV$_2$O$_4$. It is observed that the spin-orbit coupling in the
experimentally observed antiferromagnetic phase, affects the orbital order
differently in alternate V-atom chains along c-axis. This observation is found 
to be consistent with the experimental
prediction that the effect of spin-orbit coupling is intermediate between that
in case of ZnV$_2$O$_4$ and MnV$_2$O$_4$.
\end{abstract}
\vspace{0.5cm}
\maketitle

\section{INTRODUCTION}

\noindent Vanadium spinels AV$_2$O$_4$ (A=Mg, Zn, Cd) are being studied extensively
in recent years \cite{wheeler,motome,radaelli,tcherny,tm-rv,gianluca} as they provide a very
interesting playground for the study of competing interactions on a frustrated
lattice in 3-dimension. The Vanadium (V) ions at the B-sites of the spinel structure
form a pyrochlore lattice, with corner sharing tetrahedra, which is geometrically
frustrated. In its $3+$ valence state, V ion has
two electrons in the d-shell which, because of a strong Hund's coupling, align
parallel to each other thereby imparting a high spin state
($S=1$) to the ion. Thus in this family of spinels, there is a magnetic ion on
a geometrically
frustrated lattice resulting in competing ground states. Things get more
involved when the partial occupancy of triply degenerate $t_{2g}$ orbitals by
the two d-electrons makes the orbital degree of freedom unfrozen.
As both spin and orbital degrees
of freedom remain active, there is a high possibility of spin-orbit (SO) coupling
playing important role in the low energy physics of this family of systems.
Role of this interaction has been a matter of debate recently
\cite{motome,tcherny, wheeler}.
The manifestation of the interplay of orbital, spin and lattice degrees of
freedom in these systems culminates in experiments as a sequence of phase
transitions\cite{wheeler, reehuis, garlea}. A structural transition, often followed by magnetic transition
as the temperature is lowered, signifies competing interactions trying to
stabilize a particular ground state with gradual lifting of the frustration.

MgV$_2$O$_4$, with a normal spinel structure, has been reported to undergo a
structural transition at 51 K from cubic to tetragonal phase and a
magnetic transition at 42 K from non-magnetic to an antiferromagnetic(AFM) phase consisting of alternating
antiferromagnetic chains of V atoms running parallel to {\bf a} and {\bf b} directions
 as one goes along {\bf c}-axis\cite{wheeler,mamiya}. The high
temperature (HT) phase
has a cubic spinel structure with $F\bar{4}3m$ symmetry where the V ion
is surrounded by an almost perfect $O_6$ octahedron with all the six V-O bonds
having same length. This leads to a sizable ($\sim2.5$ eV) 
$t_{2g}$ - $e_{g}$ crystal field splitting of the d levels.
There is of course a small
trigonal distortion also present in this phase. Experimental results further
reveal that the structural transition to
the tetragonal phase at 51 K  is accompanied by a compression along c-axis with c/a=0.9941. This
lowers the symmetry to space
group $I\bar{4}m2$. Hence, in addition to
the $t_{2g}$ - $e_{g}$  splitting arising from roughly $O_6$ octahedral
coordination, a further splitting occurs due to the tetragonal compression
where the low lying $t_{2g}$ triplet splits into a lower energy singlet
($d_{xy}$ orbital) and a higher
energy doublet of $d_{yz}$ and $d_{zx}$ orbitals. The orbital degeneracy is thereby partially lifted with this structural distortion. Now out of the two d
electrons, one goes to the lower energy $d_{xy}$
orbital while the other still has a choice as it occupies the doubly degenerate
$d_{yz}$ and $d_{zx}$ orbitals. This opens up a possibility of orbital order
in this system. Structural transition also partially lifts the frustration of
the $V-V$ bonds in the pyrochlore lattice. This then brings in the second
transition, at lower temperature of 42 K where a long
range antiferromagnetic order sets in\cite{wheeler}. Thus the presence of any
orbital order and the magnetic order observed at low temperatures in all the
Vanadium spinels are interrelated.

Several theoretical models have been
proposed in the last few years to explain the possible orbital order in Vanadium
spinels so as to be consistent with the observed antiferromagnetic order.
Among these, the model proposed by Tsunetsugu and Motome
\cite{motome} is based on strong coupling Kugel-Khomskii
Hamiltonian and predicts an orbital order where at each V site,
d$_{xy}$ orbital is occupied by one electron and the second electron occupies
either d$_{xz}$ or d$_{yz}$ orbital, alternately, along the c-axis. However,
this type
of orbital order was found to be of lower symmetry than that ($I4_1/amd$) observed
experimentally for ZnV$_2$O$_4$. In an alternative
theoretical model, Tchernychov\cite{tcherny} considered a dominant SO interaction
which then led to the proposal that the second electron would occupy
a complex orbital of type d$_{xz}$$\pm$id$_{yz}$ at each V site. This
orbital order is found to be consistent with the underlying crystal symmetry.
Also it explains the low magnetic moment per V ion observed in these systems
as a large negative orbital moment is expected from a strong SO
coupling.
These findings were also corroborated by electronic structure calculations
\cite{tm-rv} for ZnV$_2$O$_4$.

However, recent measurements on other members of the Vanadium spinel family
raise doubts about the presence of a strong spin-orbit interaction effect.
In fact, there has been a tremendous effort, from both
theoreticians and experimentalists
working on these systems, to bring out a unified picture in terms of the
important interactions which underlie the two phase transitions (one
structural and the other magnetic). In
ZnV$_2$O$_4$ the SO coupling is found to be significant both from theory as
well as experiments\cite{tcherny,tm-rv, cheong} whereas in case of MnV$_2$O$_4$
there seems to be very little or no effect of the SO interaction on the orbital order
\cite{tm-tsd,baek}. Recently Wheeler et al.\cite{wheeler} performed neutron diffraction
measurements on MgV$_2$O$_4$
single crystal and speculated on the
basis of their observations that MgV$_2$O$_4$ might come intermediate between
ZnV$_2$O$_4$ and MnV$_2$O$_4$ as far as strength of SO coupling is concerned. Hence it is expected that in MgV$_2$O$_4$ the
occupied orbitals,
instead of being completely real (Tsunetsugu and Motome model) or completely
complex (Tchernychov model), could be a mixture of real and complex orbitals.
In the previous theoretical study on MgV$_2$O$_4$\cite{pandey}, the issue of impact of SO coupling on orbital order has not been investigated. However, as stated above, SO coupling in MgV$_2$O$_4$ is
expected to be non-negligible from experimental observations.
In order to investigate thoroughly the effect of SO interaction on magnetic and orbital
order in MgV$_2$O$_4$, we have carried out
first principle electronic structure calculations incorporating SO coupling.
Such a calculation is definitely expected to unfurl the strength
of SO coupling in this system, the nature of orbital order (if there is any)
and correlation of experimentally observed magnetic order with the orbital
order, if present.

\section{METHODOLOGY}

\noindent We undertake an electronic structure calculation
using full-potential linearized augmented-plane-wave method with the basis chosen to be linearized augmented plane waves as
implemented in WIEN2K code\cite{wien2k}.
The calculations have
been carried out with no shape approximation to the potential and charge
density. These calculations were performed at
three levels of sophistication using local spin density approximation (LSDA),
Coulomb correlated LSDA+U approximation, and with SO interaction i.e. LSDA+U+SO approximation.
To remove the self-Coulomb and self-exchange-correlation energy included in
LSDA approximation, we use self-interaction corrected scheme (LSDA+U(SIC))
\cite{anisimov},
which is appropriate for the strongly correlated systems. The corrected
energy functional is written as\cite{anisimov}
\begin{equation}\nonumber
E= E^{LSDA}
  -[UN(N-1)/2-JN(N-2)/4] + 1/2\sum_{m,m',\sigma} U_{mm'}n_{m\sigma}n_{m'-\sigma} + 1/2\sum_{m\neq m',m',\sigma}(U_{mm'}-J_{mm'})n_{m\sigma}n_{m'\sigma}
\end{equation}

\noindent Here E$^{LSDA}$ is the standard LSDA energy functional, U represents the on-site Coulomb interaction, J is the exchange parameter and
n$_{m\sigma}$ are the occupations of the localized orbitals. N is the total
number of localized electrons.

In the LSDA+U+SO calculations, SO coupling was considered within the scalar
relativistic approximation and the second variational method was employed
\cite{koelling}. In this method, the eigen value problem
is first solved separately for spin up and spin down states without
inclusion of the SO interaction term (H$_{SO}$) in the total
Hamiltonian. The resulting eigen values
and eigen functions are then used to solve new eigen value problem with the
H$_{SO}$ term in the total Hamiltonian. This method is more efficient and
computationally less expensive than the
calculation in which H$_{SO}$ is included in the total Hamiltonian by doubling the
dimension of the original eigenvalue problem in order to calculate the
non-zero matrix elements between spin-up and spin-down states. In this method,
the calculation of H$_{SO}$ matrix elements involves much less number of
basis functions than in the original basis set.

\begin{figure}
\includegraphics[width=8cm]{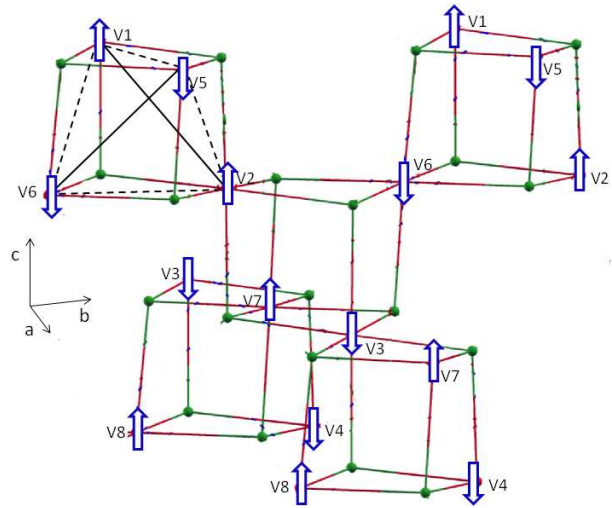}
\caption{ Corner sharing network of V$_4$O$_4$ cubes in the low temperature
structure of MgV$_2$O$_4$ showing the experimentally observed magnetic order.
The solid and dotted lines joining the V atoms (shown in one cube) represent
the shorter V-V FM bonds (2.971\AA) and longer V-V AFM bonds (2.98 \AA)
respectively. }
\end{figure}

MgV$_2$O$_4$ crystallizes in tetragonal structure with symmetry $I\bar{4}m2$
(space group 119) at low temperatures\cite{wheeler}. Atomic positions and lattice constants were taken from the
experimental data\cite{wheeler}. The atomic sphere radii were chosen to be
1.96, 1.99, and 1.78 a.u. for Mg, V, and O, respectively. We have used
50 {\bf k} points in the irreducible part of the Brillouin zone for the self-consistent calculations. In order to model
the low temperature magnetic order observed in the experiment, we have
constructed a supercell (with 8 inequivalent  Vanadium atoms). The lowering of
symmetry of this unit cell arises due to the experimentally observed antiferromagnetic ordering.
The network of corner sharing
V$_4$O$_4$ cubes of low temperature structure is shown in Fig. 1 with the magnetic order. The 8 inequivalent  Vanadium atoms considered in the calculation are also
marked in the figure with the corresponding
orientation of spins at that particular site. One can see the antiferromagnetic
chains along {\bf a} $(...V3-V7-V3-V7...)$  and {\bf b} $(...V6-V2-V6-V2...)$ axes alternating along c-axis.
 In each V$_4$O$_4$ cube there are 4 inequivalent
V atoms. Due to the presence of cooperative trigonal distortion
along c-axis resulting in alternating compression and expansion of cube
faces, there is a further symmetry breaking and hence successive cubes along
c-axis no longer remain equivalent. Furthermore, the V$_4$O$_4$ cube containing
V1, V5, V2 and V6 does not have the same spin arrangement as that containing
V3, V7, V4 and V8. Therefore to model the experimentally observed magnetic order one needs to consider 8 inequivalent V atoms in the unit
cell.

\section{RESULTS}

\begin{figure}
\includegraphics[width=9cm]{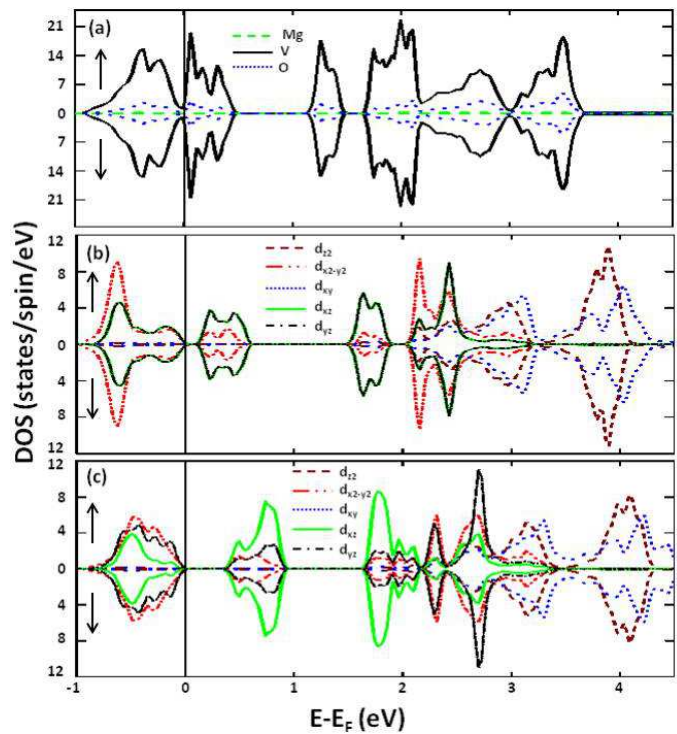}
\caption{ Spin polarized (a) total DOS within LSDA, 
(b) partial DOS for V d-states around the Fermi level
within LSDA+U, (c) partial d-DOS within LSDA+U+SO (U-J = 2 eV)in the low temperature AFM phase.}
\end{figure}

\noindent Our LSDA calculations of experimentally observed antiferromagnetically
ordered
phase show that total energy of this phase is indeed lower than that of the
corresponding
ferromagnetic (FM) phase by 0.4 eV per formula unit. The density of states (DOS) of
this antiferromagnetic
state (within LSDA) is shown in Fig. 2(a). It is observed that LSDA
gives a metallic state whereas the system is known to be a Mott insulator\cite{mamiya}.
Thus AFM interaction alone is not able to open up the gap. Around the Fermi
level, mainly V d-states are seen to be present. In an effort to have the
insulating gap as observed experimentally, we included Coulomb correlation
in our calculations within LSDA+U approximation. We performed
calculations with U$_{eff}$ (=U-J) values in the range 1 to 4 eV as found to be relevant
from the literature on
Vanadium spinel systems\cite{tm-rv,
tm-tsd,pandey}. We present the results for U$_{eff}$ = 2 eV in the
following, nevertheless it may be noted that our conclusions remain valid in the whole range of U$_{eff}$  values considered by us.
In Fig. 2(b) we show
 the partial DOS of five d-orbitals as these are the states present
around the Fermi level. As expected, the application of Coulomb correlation $U$ is able to open up
a small gap of 0.12 eV which increases with the increase in $U$. The
gap originates because of the
the splitting of the t$_{2g}$ levels in addition to the t$_{2g}$-$e_g$
splitting due to octahedral field.
The further splitting of t$_{2g}$ is
primarily caused by the antiferromagnetic interactions which
get enhanced in the presence of Coulomb correlations.

Another observation that can clearly be made from the partial DOS of d-orbitals (Fig. 2(b)) is that among the occupied t$_{2g}$ orbitals, one orbital
(i.e. d$_{x^2-y^2}$) is more populated while the other two (d$_{xz}$ and d$_{yz}$)
essentially have the same occupancy and seem to be degenerate. The higher
occupancy of d$_{x^2-y^2}$ orbital is a result of the presence of
the tetragonal compression along c-axis at low temperatures. However,
closer analysis of occupancies of the apparently degenerate d$_{xz}$ and
d$_{yz}$ orbitals at each Vanadium site shows that there is a tendency towards
orbital ordering. Table I lists the orbital occupancies of
d$_{x^2-y^2}$, d$_{xz}$ and d$_{yz}$ for the 8 inequivalent V atoms
in the unit cell considered. One clearly observes that the occupancy of
d$_{xz}$ and d$_{yz}$
orbitals are different as one moves along the c-axis whereas that of
d$_{x^2-y^2}$ remains the same. The orbital polarization increases
on increasing the value of $U$ and alternates for the d$_{xz}$ and d$_{yz}$ orbitals
(see for example, V1 and V2) in successive Vanadium layers along
c-axis. This is similar to an A-type antiferro-orbital order where the
antiferromagnetic V chains
parallel to the ab-plane have {\it ferro-orbital} order (e.g. similar orbital occupancies of V1 and V5 ions or that of
V2 and V6 ions) whereas along c-axis there is an {\it antiferro-orbital}
order between d$_{xz}$ and d$_{yz}$ orbitals (see occupancies of V1 and V2 or
that of V5 and V6 in Table I). This is consistent with the previous theoretical
observation on the same system\cite{pandey}. The observed intra-chain ferro-orbital
order is also consistent with experimental antiferromagnetic order as
per Goodenough-Kanamori-Anderson rules \cite{goodenough}.The orbital order described above is also revealed in
the calculated real space electron density at each V site shown in Fig. 3(a).
This orbital order was predicted by Tsunetsugu and Motome
\cite{motome} for Vanadium spinels from their calculations based on Kugel-
Khomskii model in strong coupling limit.

As mentioned earlier, the influence of spin-orbit coupling
on the magnetic and orbital order in these systems is continuously debated
but no conclusion has been reached yet. In order to investigate the effect of
spin-orbit interaction in this particular system, we also performed a
calculation with spin-orbit coupling within LSDA+U+SO approximation. The
solution obtained within LSDA+U+SO has a lower energy than that obtained within
LSDA+U by 0.095 eV per formula unit for $U_{eff}$=2eV.
The partial DOS (Fig. 2(c)) clearly shows a non-negligible impact of SO in general,
with an increased energy gap compared to that with LSDA+U.

\begin{figure}
\includegraphics[width=7cm]{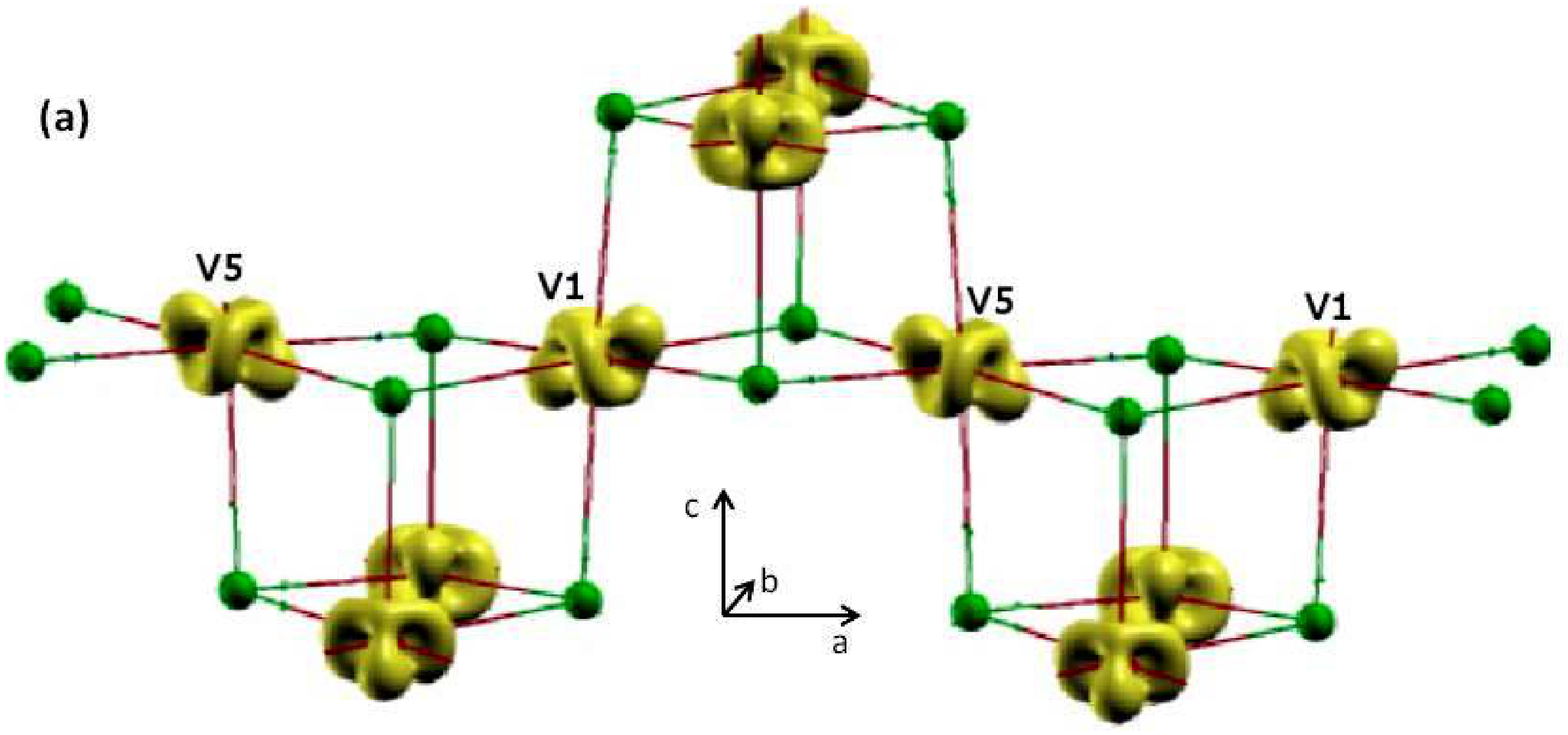}
\includegraphics[width=7cm]{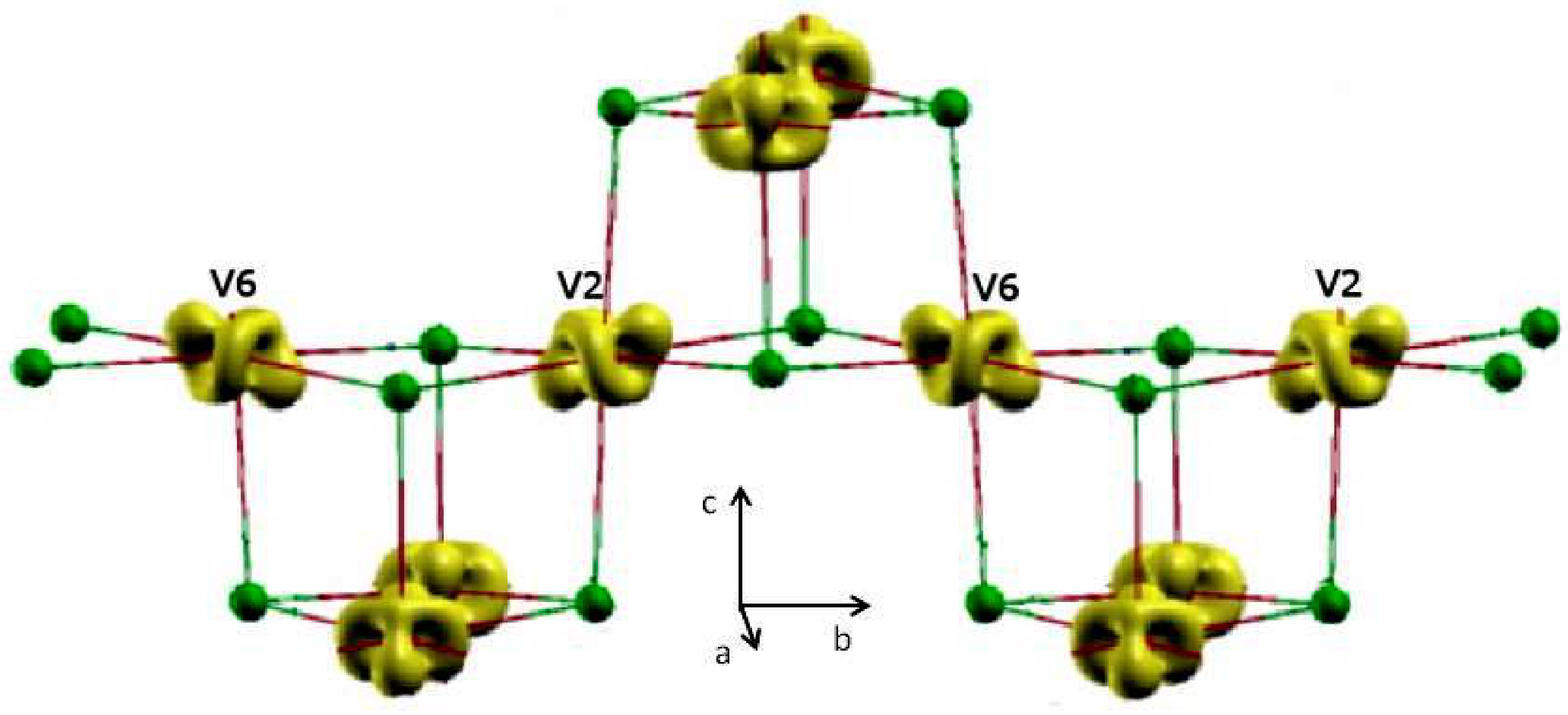}
\includegraphics[width=7cm]{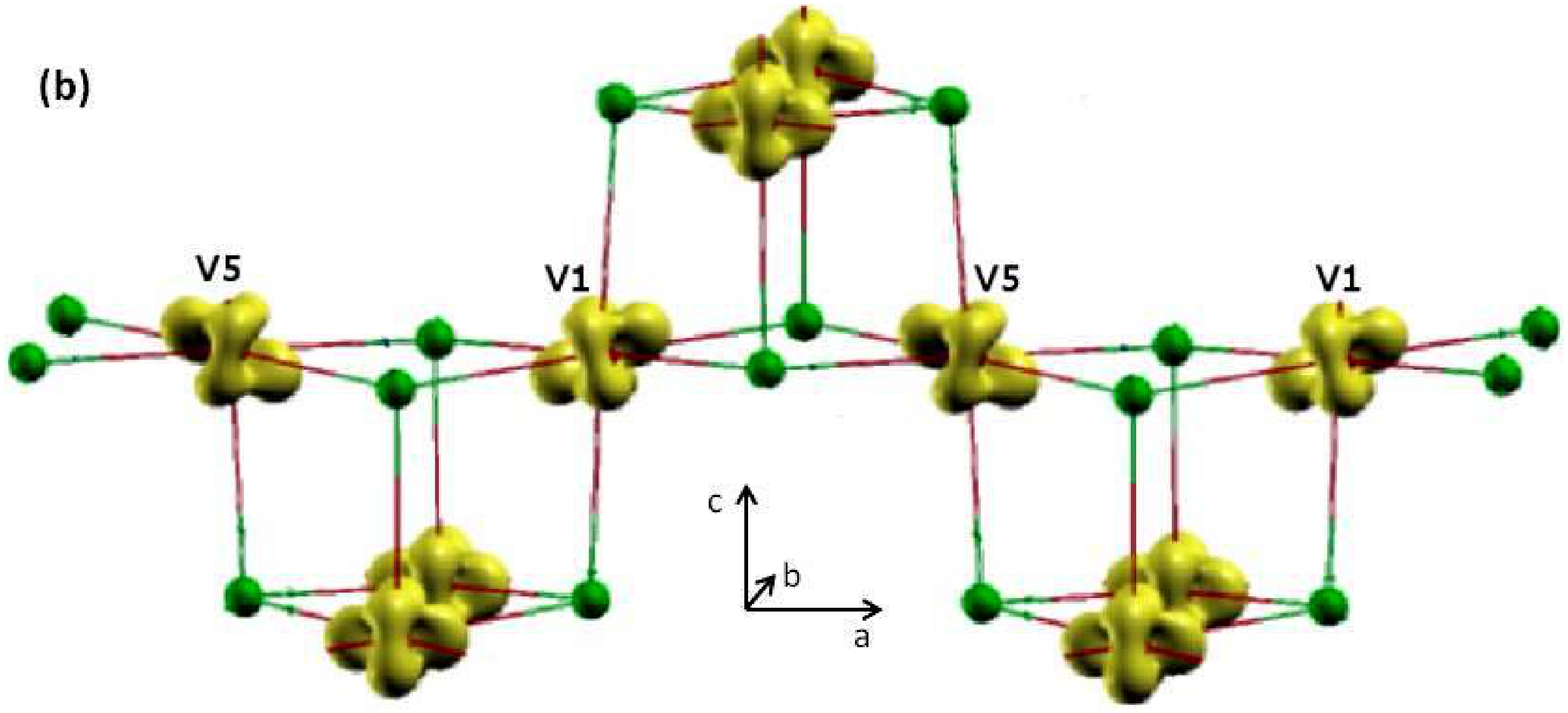}
\includegraphics[width=7cm]{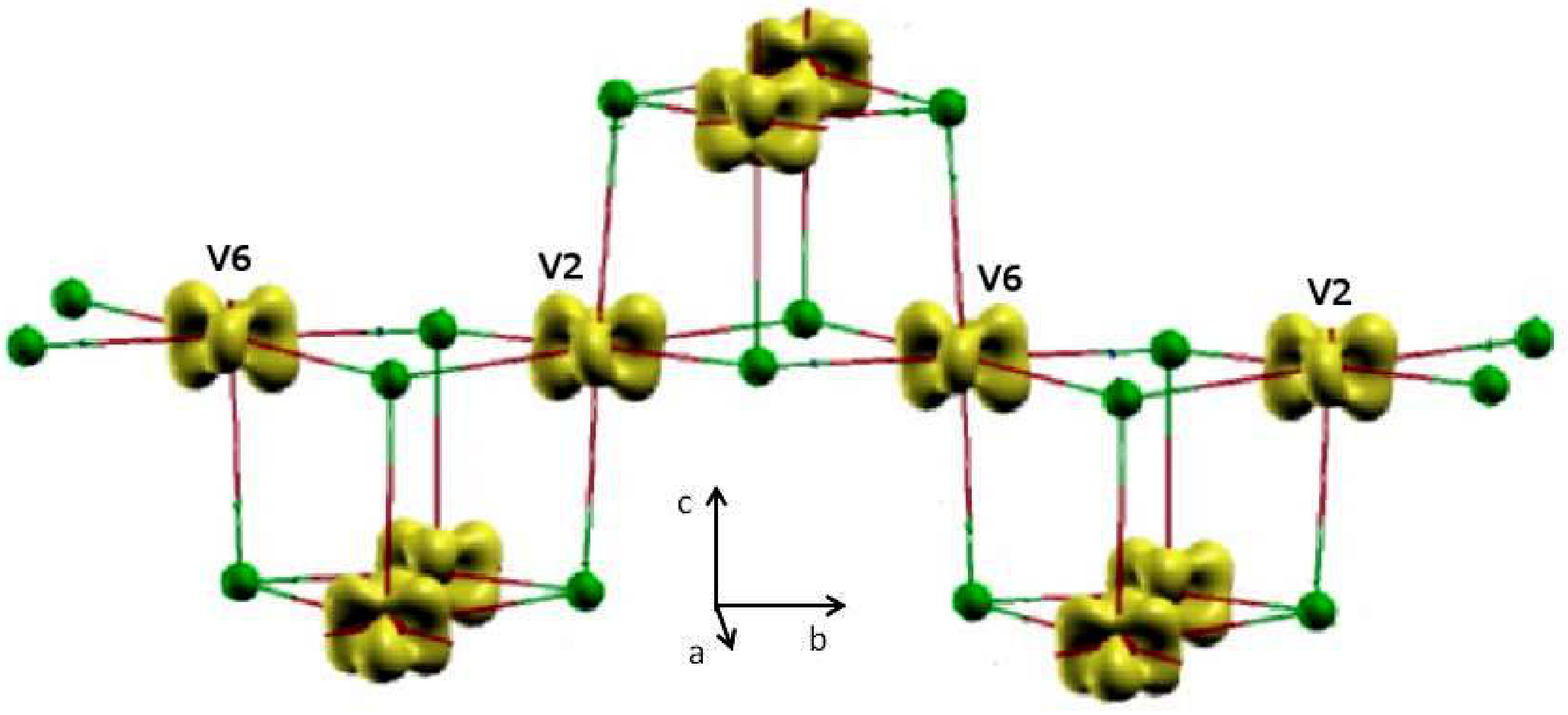}
\caption{ Real space electron density at each V site within
(a) LSDA+U (b) LSDA+U+SO along two different directions. Isosurface
used for both corresponds to 0.5 e/\AA$^3$.}
\end{figure}
The analysis of orbital occupancies in this case indeed leads to some
important and interesting observations.
The apparent degeneracy of d$_{xz}$ and d$_{yz}$
orbitals observed within LSDA+U is no longer present and there is a complete lifting of degeneracy of all the
t$_{2g}$ orbitals (see Table I). Even though, likewise LSDA+U, the antiferromagnetic V chains parallel to the
ab-plane are
still ferro-orbitally ordered and along c-axis these chains are
anti-ferro orbitally ordered, the orbital polarizations in adjacent
chains along c-axes are significantly different in the presence of SO
interaction. For example, if we
compare the occupancies (Table I) for V1 and V2 with LSDA+U and LSDA+U+SO, we note that orbital occupancies of d$_{x^2-y^2}$ are no longer
same in presence of SO interaction. Furthermore, the polarization of the d$_{xz}$ and d$_{yz}$
orbitals are also very different (i.e. at V2 the polarization of d$_{yz}$
orbital w.r.t. d$_{xz}$ is much stronger than that at V1). This implies that V
chains in successive layers along
c-axis are affected differently by the SO interaction. This is also
reflected in the orbital moments of V ions (listed in Table II and depicted
in Fig. 4). The calculated electronic density at each V site is shown
in Fig. 3(b) which brings out the impact of SO interaction. 

\begin{table}
\caption{Orbital occupancies and spin magnetic moment within  LSDA+U and LSDA+U+SO (U$_{eff}$=2 eV)}
\begin{tabular}{|c|c|c|c|c|c|}
\hline
& \multicolumn{3}{|c|}{orbital occupation} & spin magnetic\\
\cline{2-4}
 & d$_{x^2-y^2}$ & d$_{xz}$ & d$_{yz}$ & moment ($\mu_B$) \\
\hline
 \multicolumn{5}{|c|}{With LSDA+U }\\
 \hline
V1 (V3) & 0.665 & 0.553 & 0.386  & 1.53 (-1.53) \\
V2 (V4) & 0.665 & 0.386 & 0.553 & 1.53 (-1.53) \\
V5 (V7) &  0.665 & 0.553 & 0.386 & -1.53 (1.53) \\
V6 (V8) &  0.665 & 0.386 &  0.553 & -1.53 (1.53) \\ \hline
 \multicolumn{5}{|c|}{With LSDA+U+SO }\\
\hline
V1 (V3) & 0.595 & 0.451 & 0.591  & 1.574 (-1.574)  \\
V2 (V4)& 0.759 & 0.214 & 0.665 &   1.571 (-1.571)  \\
V5 (V7)& 0.595 & 0.451 & 0.591 & -1.574 (1.574)  \\
V6 (V8)&  0.759 & 0.214 & 0.665 & -1.571 (1.571)  \\ \hline
\end{tabular}
\end{table}
\begin{table}
\caption{Calculated orbital moments, total magnetic moment (J) (in $\mu_B$)
and angle of J w.r.t. z-axis within LSDA+U+SO ($U_{eff}$ = 2 eV). The spin magnetic moment is along z axis and is listed in Table I}
\begin{tabular}{|c|c|c|c|c|c|}
\hline
& \multicolumn{3}{|c|}{$\mu_{orbital}$} & $\mu_{total}$ & $angle $ \\
\cline{2-4}

 & $x$ & $y$ & $z$ & J & \\
\hline
V1 &  -0.355 & 0.000 & -0.466 & 1.15 & 17.79 \\
V2 & -0.015 & -0.030 & -0.510 & 1.05 & 1.90\\
V5 &  -0.355 & 0.000 & 0.466 & -1.15 & 162.21 \\
V6 &  -0.015 & -0.030 &  0.510  & -1.05 & 178.10 \\ \hline
\end{tabular}
\end{table}
\begin{figure}
\includegraphics[width=9cm]{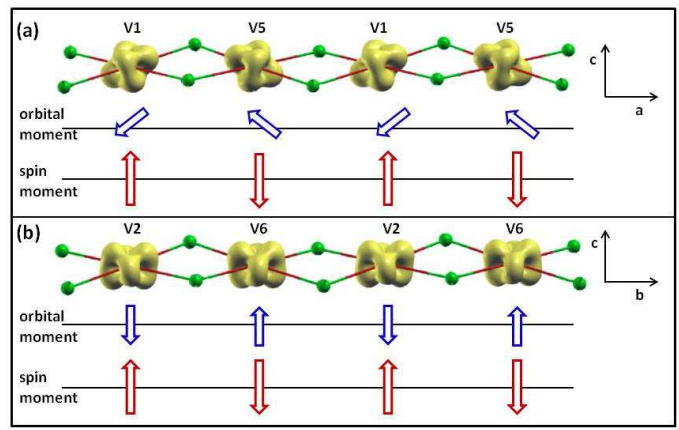}
\caption{ Electron densities for two successive Vanadium chains along c-axis within LSDA+U+SO showing the impact of SO interaction on them.
The directions of corresponding orbital and magnetic moments are also shown below
each chain.}
\end{figure}

In Fig. 4 we show two successive Vanadium chains along c-axis with
the calculated electron density at each Vanadium site in the presence
of both Coulomb correlation and SO interaction. We have also marked the
direction of orbital and magnetic moments at each site. Effect of SO
interaction is clearly different on the two chains and so is the arrangement
of the orbital moments. One chain shows the canted orbital arrangement and orbital moments are making an angle of 17.79$^{0}$ with the c-axis whereas in other chain orbital moment makes an angle of 1.90$^{0}$ (almost collinear orbital arrangement) with the c-axis (Table II). On one chain (V1-V5-V1-V5) due to canting of orbital
moment, the effect of SO interaction reduces whereas in the other chain
orbital moments align almost opposite to the magnetic moment implying
a substantial SO interaction.
The observation that the SO interaction appears to affect alternate V chains
along c-axis differently, is interesting. This also substantiates the
speculation of Wheeler et al. \cite{wheeler} that SO interaction in
MgV$_2$O$_4$ may not be as large as that in ZnV$_2$O$_4$ or as small as
that in MnV$_2$O$_4$ as discussed earlier.
 The magnitude of orbital moments observed in our calculation also corroborates this fact.
Thus our results show that a small but non-negligible
spin-orbit coupling, along with the significant trigonal distortion present in
MgV$_2$O$_4$ structure, has a substantial effect on the orbital order of this system.
This observation is consistent with the experimental observations by
Wheeler et al.\cite{wheeler} of antiferromagnetic chains with a strongly
reduced moment and the one-dimensional
behavior and a single band of excitations projected by the inelastic neutron scattering.

\section{CONCLUSIONS}

\noindent To conclude, we
have studied the effect of spin-orbit interaction on magnetic
and orbital order in the low temperature tetragonal phase of MgV$_2$O$_4$. We
observe that even
though the orbital moments are relatively small compared to those of
ZnV$_2$O$_4$, the orbital order in successive Vanadium chains is differently
affected in the presence of SO interaction. In one chain (V1-V5-V1-V5, parallel to crystallographic {\bf a} axis) the three t$_{2g}$
orbitals are nearly equally populated giving rise to a canted (non-collinear)
arrangement
of orbital moments whereas in the other (V2-V6-V2-V6, parallel to {\bf b} axis), the orbitals
are highly polarized leading to a collinear arrangement of orbital
moments. These results imply that SO
interaction in  MgV$_2$O$_4$ is non-negligible and has a significant effect on orbital order. However it is not very strong unlike ZnV$_2$O$_4$ and at the same time not
very weak unlike MnV$_2$O$_4$.

\section{ACKNOWLEDGEMENT}

\noindent This work is supported by the DST (India) fast track project (grant no.: SR/FTP/PS-74/2008). RD acknowledges CSIR (India) for a research fellowship.


\begin{thebibliography}{99}

\bibitem{wheeler} E.M. Wheeler, B. Lake, A.T. M. Nazmul Islam, M. Reehuis, P. Steffens, T. Guidi and A. H. Hill, Phys. Rev. B {\bf 82}, 140406(R) (2010).

\bibitem{radaelli} Paolo G Radaelli, New J. Phys. {\bf 7}, 53 (2005).

\bibitem{motome} H. Tsunetsugu and  Y. Motome, Phys. Rev. B {\bf 68}
060405 (2003); {\it ibid.} Prog. Theor. Phys. Suppl. {\bf 160}, 203 (2005).

\bibitem{tcherny} O. Tchernyshyov, Phys. Rev. Lett. {\bf 93} 157206 (2004).

\bibitem{tm-rv} T. Maitra and R. Valent\'{\i}; Phys. Rev. Lett. {\bf 99}, 126401, (2007).

\bibitem{gianluca} G. Giovanetti, A. Stroppa,S. Picozzi,D. Baldomir,V. Pardo,S. Blanco-Canosa,F. Rivadulla,S. Jodlauk,D. Niermann,J. Rohrkamp,T. Lorenz,S. Streltsov,D. I. Khomskii,and J. Hemberger; Phys. Rev. B {\bf 83}, 060402(R) (2011).

\bibitem{reehuis} M. Reehuis, A. Krimmel, N. Bottgen , A. Loidl and
A. Prokofiev, Eur. Phys. J. B {\bf 35}, 311 (2003).

\bibitem{garlea} V. O. Garlea, R. Jin, D. Mandrus, B. Roessli, Q. Huang, M. Miller, A. J. Schultz, and S. E. Nagler, Phys. Rev. Lett. {\bf 100}, 066404 (2008).
\bibitem{mamiya} H. Mamiya, M. Onoda, T. Furubayashi, J. Tang, and I. Nakatani,
J. Appl. Phys. {\bf 81}, 5289 (1997).

\bibitem{cheong}S. H. Jung, J. Noh1, J. Kim, C. L. Zhang, S. W. Cheong and E. J. Choi; J. Phys.: Condens. Matter {\bf 20} 175205
(2008).

\bibitem{tm-tsd} S. Sarkar, T. Maitra, Roser Valentí, and T. Saha-Dasgupta;
Phys. Rev. Lett. {\bf 102}, 216405 (2009).
\bibitem{baek} S.-H. Baek, N. J. Curro, K.-Y. Choi, A. P. Reyes, P. L. Kuhns, H. D. Zhou, and C. R. Wiebe, Phys. Rev. B {\bf 80}, 140406(R) (2009).
\bibitem{pandey}S. K. Pandey; Phys. Rev. B. {\bf 84}, 094407 (2011).
\bibitem{wien2k} P. Blaha, K. Schwartz, G. K. H. Madsen, D. Kvasnicka
and J. Luitz; WIEN2K edited by K. Schwarz (Techn. University Wien, Austria,
2001), ISBN 3-9501031-1-2.

\bibitem{anisimov} V.I. Anisimov, I.V. Solovyev, M.A. Korotin, M.T. Czyzyk, and G.A. Sawatzky, Phys. Rev. B {\bf 48}, 16929 (1993)
\bibitem{koelling}
D.D. Koelling et al., J. Phys. C 10, 3107 (1977); A.H. MacDonald et al., ibid. 13, 2675 (1980)

\bibitem{mamiya} H. Mamiya, M. Onoda, Solid State Communications {\bf 95}, 217 (1995),

\bibitem{goodenough} J. Kanamori, Prog. Theor. Phys. {\bf 17} 177 (1957); J. B.
Goodenough, J. Phys. Chem. Solids {\bf 6}, 287 (1958); P. W. Anderson, Phys.
Rev. {\bf 79} 350 (1950).




\end{thebibliography}
\end{document}